\documentclass[twocolumn,showpacs,aps,prb,reprint]{revtex4-1}

\pdfoutput=1

\usepackage{epsfig}
\usepackage{amsmath}
\usepackage{amssymb}
\usepackage{bm}
\usepackage{color}
\def\sgn{\mathop{\textrm{sgn}}}

\newcommand{\beq}{\begin{equation}}
\newcommand{\eeq}{\end{equation}}
\newcommand{\beqarray}{\begin{eqnarray}}
\newcommand{\eeqarray}{\end{eqnarray}}
\newcommand{\eq}[1]{Eq.~(\ref{#1})} 
\newcommand{\fig}[1]{Fig.~(\ref{#1})} 
\newcommand{\Ref}[1]{Ref.~\onlinecite{#1}} 

\newcommand{\tos}[2]{\ensuremath{\hat{\tau}_{#1}\otimes\hat{\sigma}_{#2}}}

\begin{document}

\allowdisplaybreaks

\title{Topological Yu-Shiba-Rusinov chain from spin-orbit coupling} 
\author{P. M. R. Brydon}
\email{pbrydon@umd.edu}
\affiliation{Condensed Matter Theory Center and Joint Quantum
  Institute, University of Maryland, College Park, Maryland
  20742-4111, USA}
\author{S. Das Sarma}
\affiliation{Condensed Matter Theory Center and Joint Quantum
  Institute, University of Maryland, College Park, Maryland
  20742-4111, USA}
\author{Hoi-Yin Hui}
\affiliation{Condensed Matter Theory Center and Joint Quantum
  Institute, University of Maryland, College Park, Maryland
  20742-4111, USA}
\author{Jay D. Sau}
\affiliation{Condensed Matter Theory Center and Joint Quantum
  Institute, University of Maryland, College Park, Maryland
  20742-4111, USA}

\date{January 22, 2015}

\begin{abstract}
We investigate the possibility of realizing a topological state in the
impurity band formed by a chain of classical spins embedded in a
two-dimensional singlet superconductor with Rashba spin-orbit coupling. In
contrast to similar proposals which require a helical spin texture of
the impurity spins for a nontrivial topology, here we show that
spin-flip correlations due to the spin-orbit coupling in the
superconductor produces a topological state for
ferromagnetic alignment of the impurity spins. From the
Bogoliubov-de Gennes equations we derive an effective
tight-binding model for the subgap states which resembles a spinless
superconductor with long-range hopping and pairing terms. We evaluate
the topological invariant, and show that a topologically non-trivial
state is generically present in this model. 
\end{abstract}



\maketitle

\section{Introduction}

The search for a condensed-matter realization of
the Majorana fermion continues, motivated both by the underlying
fundamental physics and potential technological
applications. Such states are predicted to occur in vortices of
unconventional superconductors.~\cite{vortex} They may also be
realized as edge states
of engineered spinless superconductors with nontrivial topology, 
e.g. a Kitaev chain,~\cite{Kitaev2001} in a superconducting
heterostructure.~\cite{hetero} Such a
phase was predicted to occur in a semiconducting
nanowire in proximity contact with a superconductor and in an applied
magnetic field.~\cite{Lutchyn2010,Oreg2010} Transport signatures
consistent with this theoretical prediction were subsequently
detected,~\cite{Mourik2012,Das2012} although the definitive existence of the
Majorana mode in such devices is still debated.~\cite{nanowireprob} 

Much attention has recently been directed at an alternative proposal,
where a topological band arises from the overlapping Yu-Shiba-Rusinov (YSR)
states~\cite{Shiba} in a 
chain of magnetic impurities with helical spin order on the surface of
a superconductor.~\cite{Choy2011,NadjPerge2013,Nakosai2013,Klinovaja2013,Braunecker2013,Vazifeh2013,Pientka2013,Poeyhoenen2014,Pientka2014,Reis2014,Roentynen2014,NadjPergeAPS2014}
The helical spin texture plays a critical role combining the
  effect of the spin-orbit coupling (SOC) and external field in the 
nanowire proposal. Topological states are similarly predicted in
metallic systems with coexisting superconductivity and helical magnetic
order.~\cite{Kjaergaard2012,Martin2012,Kotetes2013} A significant
advantage of the YSR chain proposal is that it
is possible to unambiguously image the Majorana end modes using
scanning tunneling microscopy (STM), in contrast to relying on
difficult-to-interpret transport measurements of nanowire systems. 
Although critical to ensuring a topological state, the helical order
also represents the main experimental difficulty since it is impossible
to control externally. The helical order is stable when the
magnetic ions are placed on a quasi-one-dimensional
substrate,~\cite{Klinovaja2013,Braunecker2013,Vazifeh2013} but for the
physically-relevant case of a planar surface the chain is generically  
unstable towards a ferromagnetic or antiferromagnetic
configuration.~\cite{Kim2014} A 
pair-breaking effect in the superconducting state might nevertheless
restore the stability of the helical order,~\cite{Reis2014} but
disorder effects may still turn out to be a strong detrimental
factor.~\cite{Kim2014} 

The prospect of unambiguously verifying the existence of
Majorana end modes in a YSR chain motivates the search for a way
to realize a topological state in this system without relying upon
an intrinsic helical ordering of the impurity 
spins. For example, it has been proposed to use external magnetic
fields and a supercurrent flow to tune a nontopological
antiferromagnetic chain into a topological
regime.~\cite{Heimes2014} In view of the importance of SOC in the
nanowire proposal, it is also interesting to include SOC in the 
description of the superconducting host of the impurity chain. Indeed,
one typically expects the presence of a Rashba SOC
at the surface of the superconductor due to the broken inversion
symmetry. We note that SOC intrinsic to the
superconductor has been considered in other proposals for realizing
topological systems,~\cite{SOCformajorana} and the possible relevance
of SOC in the context of a topologically nontrivial magnetic impurity
chains has been mentioned in~\Ref{Kim2014}. This
scenario has recently been invoked to explain STM measurements of
zero-bias peaks at the ends of a 
ferromagnetic chain on a superconductor,~\cite{Yazdani} although
the relevance of YSR physics to this situation is uncertain, as we
discuss below. In spite of the  
extensive activity on the interplay between magnetic chains and
superconductivity in generating emergent topological
phases,~\cite{Choy2011,NadjPerge2013,Nakosai2013,Klinovaja2013,Braunecker2013,Vazifeh2013,Pientka2013,Poeyhoenen2014,Pientka2014,Reis2014,Roentynen2014,NadjPergeAPS2014,Heimes2014} 
the specific problem of combining both spin-orbit coupling and
magnetic YSR chain physics together in a model has been conspicuously
lacking. We study this
particular issue in our current work by 
generalizing and synthesizing the existing work in the literature,
most specifically
Refs.~\onlinecite{Choy2011},~\onlinecite{Pientka2013},
and~\onlinecite{Kim2014}.  

We mention here that very recently there has been a spurt in the
activity~\cite{NadjPerge2014,Hui2014,Li2014,Dumitrescu2014}
on topological superconductivity and 
emergent Majorana fermions, following Ref.~\onlinecite{Yazdani}, in
ferromagnetic chains 
fabricated on the surface of bulk superconductors. In particular, a
report of impressive STM
experiments~\cite{NadjPerge2014} has just appeared claiming the
generic observation of 
Majorana fermions at the ends of Fe chains on superconducting Pb.
These experiments are the main reason for this enhanced
activity, but the question of the topological
nature of purely ferromagnetic chains proximity coupled to $s$-wave
superconductors is of intrinsic theoretical interest 
independent of experimental developments. Rather surprisingly, the
theory of such systems has not yet been dealt with in any detail,
in contrast to the extensive theoretical analyses
of spiral YSR chains and spin-polarized semiconducting
nanowires. Although there is superficial  
similarity between the model used in the current work and the
experimental systems,~\cite{NadjPerge2014,Yazdani} it is far too early
to tell whether there is any connection between the theory and
predictions presented in the current work involving a weakly coupled
YSR chain and the experimental system involving Fe chains where
tunneling between the atoms may well be strong. The corresponding
topological theory for strongly-coupled ferromagnetic chains on
superconducting substrates has been considered
in Refs.~\onlinecite{NadjPerge2014,Hui2014,Li2014,Dumitrescu2014},
and is a generalization of previous proposals to realize Majorana
fermions in half-metals (i.e. fully spin-polarized ferromagnets)
deposited on 
superconductors.~\cite{SOCformajorana} Further discussion of such
strongly tunnel coupled ferromagnetic nanowire systems and the
experimental results of Ref.~\onlinecite{NadjPerge2014} is beyond the
scope of the current work.

In this paper we show that the SOC indeed
induces a topological state in a YSR chain formed from
ferromagnetically aligned impurity spins, and so demonstrate that the
more delicate helical order is not essential to such
proposals. To this end, we 
analytically construct a tight-binding model for the YSR states
valid in the limit of ``deep'' impurities, when the impurity band lies
close to the middle of the superconducting gap. Although the
SOC does not affect the YSR states for an isolated
impurity, it dramatically alters the results for the
chain. Specifically, spin-flip correlations in the bulk
superconductor, induced by the antisymmetric SOC, 
mix the two branches of the impurity band when the polarization of the
impurity spins is transverse to the  SOC along the
chain. This can be interpreted as a triplet pairing  
amplitude in a Kitaev-like model, and is thus responsible for the
topologically nontrivial state. A magnetic polarization parallel to
the SOC, on the other hand, produces no such mixing
but instead results in an asymmetric dispersion with trivial topology.
We construct a phase diagram, demonstrating that a topological state
is possible for infinitesimal SOC strength.
Our analysis closely follows that of~\Ref{Pientka2013}, where a
similar tight-binding model for the impurity band was obtained for a
chain with spiral magnetic texture embedded in a three-dimensional
superconductor.

\section{Model}

A bulk two-dimensional
singlet $s$-wave superconductor with Rashba SOC is described by the
Hamiltonian $H = \sum_{\bf k}\Psi^{\dagger}_{\bf
  k}\check{H}_{\bf k}\Psi^{}_{\bf k}$ where 
\beq
\check{H}_{\bf k} = \hat{\tau}_{z}\otimes(\xi_{\bf k}\hat{\sigma}_0
  + {\bf l}_{\bf k}\cdot{\hat{\pmb{\sigma}}}) +
  \Delta\tos{x}{0}\,. \label{eq:bulkH}
\eeq
Here $\hat{\tau}_{\mu}$ ($\hat{\sigma}_{\mu}$) are the Pauli matrices
in Nambu (spin) space,  and  $\Psi^{}_{\bf k} 
= (c^{}_{{\bf k},\uparrow}, c^{}_{{\bf k},\downarrow},
c^{\dagger}_{-{\bf k},\downarrow}, -c^{\dagger}_{-{\bf
    k},\uparrow})^{T}$ is the spinor of creation and annihilation
operators. We have adopted the notation that $\hat{\ldots}$ and
  $\check{\ldots}$ indicate $2\times2$ and $4\times4$ matrices,
  respectively.  The noninteracting dispersion is given by $\xi_{\bf k}
= \hbar^2{\bf k}^2/2m - \mu$ where $m$ is the effective mass and $\mu$
the chemical potential, the Rashba SOC
is parametrized by ${\bf l}_{\bf k} = \lambda(k_y{\bf e}_x - k_x{\bf e}_y) =
\lambda k(\sin\theta{\bf e}_x -\cos\theta{\bf e}_y)$ where $\lambda$
is the SOC strength, and $\Delta$ is the superconducting gap.

The SOC lifts the spin degeneracy in the normal state,
resulting in the dispersions $\xi_{{\bf k},\pm} = \xi_{\bf k} \pm |{\bf
  l}_{\bf k}|$, where the plus (minus) sign corresponds to the
positive (negative) helicity band. As time-reversal symmetry remains
intact, however, in the superconducting phase there is only pairing
between states in the same helicity band. The bulk Green's function
can then be written as $\check{G}_{\bf k}(\omega) =
\tfrac{1}{2}\{\check{G}^{+}_{\bf k}(\omega) + \check{G}^{-}_{\bf
  k}(\omega)\}$, where
\beqarray
\check{G}^{\pm}_{\bf k}(\omega) & = & \left(\omega\hat{\tau}_0 + \xi_{\pm}\hat{\tau}_{z} +
\Delta\hat{\tau}_x\right)\otimes\left(\hat{\sigma}_0 \pm \sin\theta
\hat{\sigma}_x \mp \cos\theta\hat{\sigma}_y\right) \notag \\
&& \times \left(\omega^2 - \xi_{\pm}^2 - \Delta^2\right)^{-1}\,, \label{eq:Greensfunction}
\eeqarray
is the Green's function in each helicity sector. Note that the
SOC produces normal spin-flip and triplet pairing
terms in the Green's function.~\cite{Gorkov2001} For clarity
we suppress the momentum index in the dispersion of the helical bands,
i.e. $\xi_{{\bf k},\pm} \equiv \xi_{\pm}$.

\section{Single impurity}

We first consider a single (classical)
magnetic impurity with spin ${\bf S}$ at the origin, interacting with
the electron states with exchange strength $-J$. We include this in our
model by adding $H_{\text{imp}} = -J{\bf 
  S}\cdot[\Psi^\dagger({\bf
    0})\hat{\tau}_0\otimes\hat{\pmb{\sigma}}\Psi({\bf 0})]$ to the
bulk Hamiltonian, where
$\Psi({\bf r}) = \int \frac{d^2k}{(2\pi)^2}\Psi_{\bf k}e^{i{\bf
    k}\cdot{\bf r}}$. We aim to solve the Bogoliubov-de Gennes
equation $(H + H_{\text{imp}})\psi({\bf r}) = \omega \psi({\bf r})$
for the impurity bound states, i.e. for energy $|\omega| <
\Delta$. By straightforward manipulation,~\cite{Pientka2013} the
spinor of the bound state at the impurity $\psi({\bf 0})$ satisfies
the equation 
\beq
\left\{\check{1} + \int \frac{d^2k}{(2\pi)^2}\check{G}_{\bf
  k}(\omega)J{\bf
  S}\cdot(\hat{\tau}_0\otimes\hat{\pmb{\sigma}})\right\}\psi({\bf 0})
= 0\,. \label{eq:singleimpurity1}
\eeq
To evaluate this equation, we split the Green's function into positive
and negative helicity components and then convert the integral over
the momentum to an integral over the appropriate dispersion
$\xi_{\pm}$ and the angle $\theta$  
\beq
\int\frac{d^2k}{(2\pi)^2}\check{G}^{\pm}_{\bf k}(\omega) \approx \frac{\cal
  N_{\pm}}{2\pi}\int^{D}_{-D}d\xi_{\pm} \int^{2\pi}_{0}d\theta
\check{G}^{\pm}_{\bf k}(\omega) \,, \label{eq:intapprox}
\eeq
where ${\cal N}_{\nu} = (m/2\pi\hbar^2)[1 \mp
  \widetilde{\lambda}/(1 + \widetilde{\lambda}^2)^{1/2}]$ is the
density of states of the $\nu=\pm$ 
helicity band at the Fermi level, $\widetilde{\lambda} = \lambda
m/\hbar^2 k_F$ is the ratio of SOC splitting to the Fermi energy and gives a
dimensionless measure of the SOC strength,
$k_F$ the Fermi wavevector in the absence of SOC,
and $D\rightarrow\infty$ is a
cutoff. The symmetric cutoff in~\eq{eq:intapprox} is used for
  simplicity; Although it implies particle-hole symmetry of the normal
  dispersion, relaxing this assumption does not qualitatively change
  our results. The 
resulting integrals are presented in the appendix. Due to 
the isotropic $\delta$-function structure 
of the potential, the integrals involving the spin-flip and triplet
pairing terms in the Green's function vanish,
and~\eq{eq:singleimpurity1} therefore has exactly the same form as 
a magnetic impurity in an $s$-wave superconductor without
SOC,~\cite{Shiba,Pientka2013} specifically 
\beq
\left\{\check{1} - \frac{\alpha}{\sqrt{\Delta^2 - \omega^2}}\left[\omega \hat{\tau}_{0} +
  \Delta\hat{\tau}_x\right]\otimes({\bf e}_{\bf
  S}\cdot\hat{\pmb{\sigma}})\right\}\psi({\bf 0}) = 0\,,
\eeq
where $\alpha = \tfrac{\pi}{2}({\cal N}_{+} + {\cal N}_{-})JS$, $S =
|{\bf S}|$, and ${\bf e}_{\bf S} = {\bf S}/S$. The solutions of this
equation occur at $\omega = \pm \epsilon_0$, where
$\epsilon_{0} = \Delta (1-\alpha^2)/(1+\alpha^2)$.
The form of the corresponding spinors $\psi_{\pm}({\bf 0})$ is
dictated by the orientation of the impurity spin. Parametrizing ${\bf
  S} = S(\cos\eta\sin\zeta,\sin\eta\sin\zeta,\cos\zeta)$, these
spinors can then be written~\cite{Pientka2013} up to unimportant
normalization constant as
\beq
\psi_{+}({\bf 0}) = \left(\begin{array}{c}
\chi_{\uparrow}\\ \chi_{\uparrow} \end{array}\right)\,, \qquad
\psi_{-}({\bf 0}) = \left(\begin{array}{c}
\chi_\downarrow\\ -\chi_\downarrow \end{array}\right)\,, \label{eq:Shiba} 
\eeq
where 
\beqarray
\chi_\uparrow &=& \left(\begin{array}{cc} \cos\zeta/2\,, &
  e^{i\eta}\sin\zeta/2\end{array}\right)^T\,, \\ 
\chi_\downarrow &=& \left(\begin{array}{cc}
  e^{-i\eta}\sin\zeta/2\,, & -\cos\zeta/2\end{array}\right)^T.
\eeqarray

\section{Ferromagnetic chain}

The above analysis can be extended to a chain
of ferromagnetically-aligned impurity spins, with the impurity
Hamiltonian now written as 
\beq
{\cal H}_{\text{imp}} = -J\sum_{j}{\bf S}\cdot[\Psi^\dagger({\bf
    r}_j)\hat{\tau}_0\otimes\hat{\pmb{\sigma}}\Psi({\bf r}_j)]\,,
\eeq
where ${\bf r}_{j}$ is the position of the $j$th impurity. We have
suppressed the site index of the spins since they all point in the
same direction. 
 Without loss of
generality, we assume that the chain runs along the $x$-axis, and so
${\bf r}_{j} = x_{j}{\bf e}_x$. After similar manipulations as in the
single impurity problem, the BdG equations for the subgap YSR states
on the chain can be written
\begin{multline}
\left\{\check{1} - \frac{\alpha}{\sqrt{\Delta^2 - \omega^2}}\left[\omega \hat{\tau}_{0} +
  \Delta\hat{\tau}_x\right]\otimes({\bf e}_{\bf
  S}\cdot\hat{\pmb{\sigma}})\right\}\psi(x_i) \\
=  -\sum_{j\neq i}\check{J}(x_{ij}){\bf
  e}_{\bf S}\cdot(\hat{\tau}_{0}\otimes\hat{\pmb{\sigma}})\psi(x_j) \label{eq:multiimp}
\end{multline}
where $x_{ij} = x_i - x_j$ and the matrix $\check{J}(x_{ij})$ is defined
\begin{widetext}
\beqarray
\check{J}(x_{ij}) &=& JS\int \frac{d^2k}{(2\pi)^2} \check{G}_{\bf
  k}(\omega)e^{ik_xx_{ij}} \notag \\
&=& \frac{JS}{2}\left\{[I^{-}_1(x_{ij}) +
  I^{+}_{1}(x_{ij})]\hat{\tau}_{z}\otimes\hat{\sigma}_{0} + \omega[I^{-}_3(x_{ij}) +
  I^{+}_3(x_{ij})]\hat{\tau}_{0}\otimes\hat{\sigma}_{0} + \Delta[I^{-}_3(x_{ij}) +
  I^{+}_3(x_{ij})]\hat{\tau}_{x}\otimes\hat{\sigma}_{0} \right.\notag \\
& & \left.  + [I^{-}_2(x_{ij}) -
  I^{+}_2(x_{ij})]\hat{\tau}_{z}\otimes\hat{\sigma}_{y} + \omega[I^{-}_4(x_{ij}) -
  I^{+}_4(x_{ij})]\hat{\tau}_{0}\otimes\hat{\sigma}_{y} + \Delta[I^{-}_4(x_{ij}) -
  I^{+}_{4}(x_{ij})]\hat{\tau}_{x}\otimes\hat{\sigma}_{y}
\right\}\,. \label{eq:J}
\eeqarray
\end{widetext}
We have expressed this in terms of the integrals
\begin{subequations} \label{eq:impintegrals}
\beqarray
I^{\nu}_{1}(x) & = &  \frac{{\cal N}_\nu}{2\pi}\int^{D}_{-D}
d\xi \int^{2\pi}_{0}d\theta \frac{\xi e^{ik_{\nu}(\xi)x\cos\theta}}{\omega^2 - \xi^2 -
  \Delta^2} \,, \\
I^{\nu}_{2}(x) & = & \frac{{\cal N}_\nu}{2\pi}\int^{D}_{-D}
d\xi \int^{2\pi}_{0}d\theta \frac{\xi e^{i\theta} e^{ik_{\nu}(\xi)x\cos\theta}}{\omega^2 - \xi^2 -
  \Delta^2} \,, \\
I^{\nu}_{3}(x) & = & \frac{{\cal N}_\nu}{2\pi}\int^{D}_{-D}
d\xi \int^{2\pi}_{0}d\theta \frac{e^{ik_{\nu}(\xi)x\cos\theta}}{\omega^2 - \xi^2 -
  \Delta^2}\,, \\
I^{\nu}_{4}(x) & = & \frac{{\cal N}_\nu}{2\pi}\int^{D}_{-D}
d\xi \int^{2\pi}_{0}d\theta \frac{e^{i\theta}
  e^{ik_{\nu}(\xi)x\cos\theta}}{\omega^2 - \xi^2 - 
  \Delta^2} \,,
\eeqarray
\end{subequations}
where $k_\nu(\xi) = k_{F,\nu} + \xi/\hbar v_{F,\nu}$, while $k_{F,\nu} =
k_F[(1+\widetilde{\lambda}^2)^{1/2} - \nu\widetilde{\lambda}]$
and $v_{F,\nu}= (\hbar k_F/m)(1+\widetilde{\lambda}^2)^{1/2}$ are the Fermi vector and velocity for the $\nu$
helicity band, respectively. These integrals are explicitly evaluated
in the appendix for $D\rightarrow\infty$, where we
also provide asymptotic 
expansions valid for $k_{F,\nu}|x|\gg 1$. Note that $I^{\nu}_1(x)$ and
$I^{\nu}_3(x)$ are even functions of $x$, whereas $I^{\nu}_2(x)$ and
$I^{\nu}_4(x)$ are odd. 

In contrast to the single-impurity system considered above, the
presence of SOC makes a significant difference to 
the BdG equations for the multi-impurity problem: while the first line
of~\eq{eq:J} is identical to the result found in~\Ref{Pientka2013},
the second line is only present for nonzero SOC. This
line contains explicitly magnetic terms $\propto \hat{\sigma}_{y}$,
reflecting the orientation of the SOC vector ${\bf
  l}_{\bf k} || {\bf e}_y$ for ${\bf k}$ pointing along the magnetic
chain.

\section{Tight-binding model}

We do not attempt a general solution
of~\eq{eq:multiimp}, but instead consider the analytically-tractable
limit of dilute ``deep'' impurities, as discussed
in~\Ref{Pientka2013}. Specifically, we assume that $\alpha \approx 1$,
so that the energy $\epsilon_0$ of the isolated YSR state lies close to the
center of the gap, and that the spacing $a$ between impurities
is sufficiently large that the impurity band formed from the
hybridized YSR states lies entirely within the superconducting
gap. Linearizing the BdG equations~\eq{eq:multiimp} in the
energy $\omega$ and the coupling between impurity sites, we obtain
after straightforward manipulation
\begin{widetext}
\beq
\Delta\left[{\bf e}_{{\bf
  S}}\cdot(\hat{\tau}_0\otimes\hat{\pmb{\sigma}}) -
\alpha\hat{\tau}_x\otimes\hat{\sigma}_{0}\right]\psi(x_i) + \Delta\sum_{j\neq i}{\bf e}_{{\bf
  S}}\cdot(\hat{\tau}_0\otimes\hat{\pmb{\sigma}})\lim_{\omega\rightarrow0}\check{J}(x_{ij}){\bf
  e}_{\bf S}\cdot(\hat{\tau}_{0}\otimes\hat{\pmb{\sigma}})\psi(x_j) =
\omega \psi(x_i)
\eeq
\end{widetext}
This equation is now projected into the YSR states [\eq{eq:Shiba}]
at each site, to obtain a BdG-type equation for the impurity band
\beq
\widetilde{H}(i,j)\phi_j = \omega \phi_i
\eeq 
where $\phi_i = (u_{i,+},u_{i,-})^{T}$ is the vector of the
wavefunctions for the $+$ and $-$ YSR states at site $i$ and
\beq
\widetilde{H}(i,j) = \left(\begin{array}{cc}
A_{ij} + B_{ij} & C_{ij} \\ C^\ast_{ji} & -A_{ij} + B_{ij} \end{array}\right) \label{eq:Heffrealspace}
\eeq
where
\beqarray
A_{ij} & = & \epsilon_{0}\delta_{ij} +
\frac{1}{2}JS\Delta^2\lim_{\omega\rightarrow0}\left[I^{+}_{3}(x_{ij}) +
  I^{-}_{3}(x_{ij})\right] \,,\\
B_{ij} & = &  \frac{1}{2}JS\Delta^2\sin\eta\sin\zeta\lim_{\omega\rightarrow0}\left[I^{-}_{4}(x_{ij}) -
  I^{+}_{4}(x_{ij})\right] \,,\\
C_{ij} & = & 
-\frac{i}{2}JS\Delta\left(\cos^2\tfrac{\zeta}{2} +
  \sin^2\tfrac{\zeta}{2}e^{-2i\eta}\right)\notag \\
&& \times\lim_{\omega\rightarrow0}\left[I^{-}_{2}(x_{ij}) -
  I^{+}_{2}(x_{ij})\right] \,.
\eeqarray
Note that the integrals in these expressions are to be regarded as
vanishing for $i = j$.

The effective tight-binding Hamiltonian~\eq{eq:Heffrealspace} is the
central result of this paper. Due to the antisymmetry of the integrals
$I^{\nu}_{2}(x)$ in the off-diagonal terms, it can be interpreted as
describing superconducting spinless fermions, recalling the 
Kitaev model,~\cite{Kitaev2001} albeit with long-range hopping and
pairing terms. The properties of this system depend
crucially on the SOC in the bulk superconductor and
the polarization of the impurity spins. Specifically, the pairing term
$C_{ij}$ is only present for nonvanishing
SOC, and when the polarization of the ferromagnetic
chain has a component perpendicular to the $y$-axis. Examining~\eq{eq:J},
we observe that the pairing term
originates from the spin-flip correlations in the host superconductor 
induced by the SOC. A polarization
component along the $y$-axis contributes an antisymmetric hopping
$B_{ij}$ in the presence of SOC. This echoes the
asymmetric dispersion of a spin-orbit coupled electron gas in the
direction of an applied magnetic field, and its appearance here is due
to the triplet pairing correlations in the bulk Green's
function~\eq{eq:Greensfunction}.  

A similar tight-binding model was derived in~\Ref{Pientka2013},
but there the odd-parity pairing term arose from the spiral magnetic
texture of the impurity chain. This mechanism for generating a pairing
term is still valid in the presence of the SOC
considered here. Examining the interplay of spiral spin texture and
SOC is an interesting topic which we leave to later work.

\section{Topological properties}

To conclude we examine the topology
of the impurity band. For an infinite chain with uniform spacing $a$ of
the impurities, we define the Fourier transform of the
Hamiltonian~\eq{eq:Heffrealspace} 
\beq
\widetilde{H}(k) = \left(\begin{array}{cc}
A(k) + B(k) & C(k) \\
C^\ast(k) & -A(k) + B(k)\end{array}\right) \label{eq:Heffmomentum}
\eeq
where $A(k) = \sum_{j}A_{0j}e^{ikja}$, etc. Using the asymptotic
forms for the integrals, it is possible to obtain analytical
expressions for these quantities in the limit $k_{F,\nu}a\gg1$, which
are presented in the appendix. The
Hamiltonian~\eq{eq:Heffmomentum} is in Altland-Zirnbauer symmetry
class D, and for a fully-gapped system it is therefore characterized
by the $\mathbb{Z}_2$ topological invariant~\cite{Kitaev2001}
\beq
Q = \sgn\{A(0)A(\pi/a)\}\,.
\eeq
The system is topologically nontrivial for $Q=-1$; conversely, $Q=1$
indicates a trivial state. 

\begin{figure}
\includegraphics[width=\columnwidth]{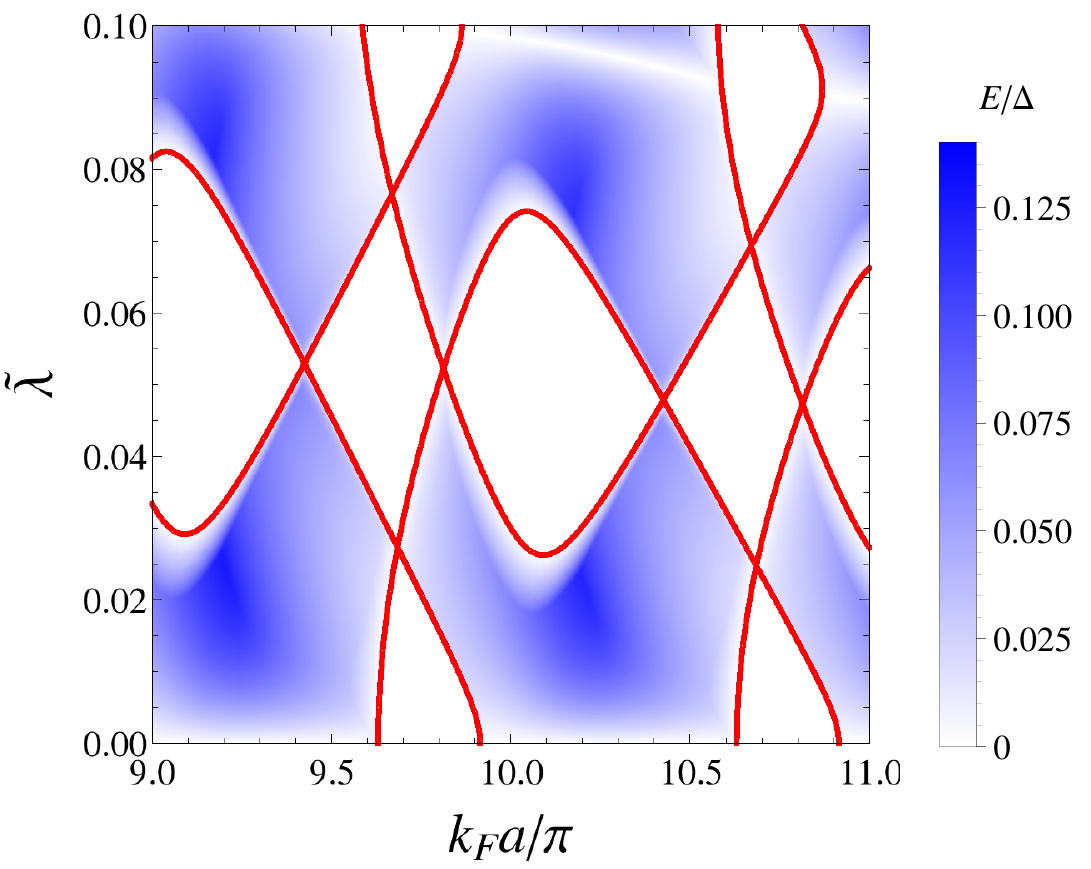}
\caption{(Color online). Topological phase diagram for the effective
  model as a function of $k_Fa$ and $\widetilde{\lambda}$. The
  topological regions are shaded according to the magnitude of the
  gap, while the nontopological regions are left blank. Red lines
  indicate the boundary between topological and nontopological
  phases. We have chosen
  $\epsilon_0=0$ for the isolated impurity level and $\xi_0 = 5a$ for
  the superconducting coherence length at
  $\widetilde{\lambda}=0$, which ensures that the impurity band remains
  within the superconducting gap. The impurity spins point in the
  $x$-$z$ plane. The
  large values of $k_Fa$ allow us to utilize the asymptotic
  expressions for the entries in~\eq{eq:Heffmomentum}.
}\label{phasediagram} 
\end{figure}

To demonstrate that our model supports a topologically nontrivial
state, in~\fig{phasediagram} we present a phase diagram as a function of the
dimensionless SOC $\widetilde{\lambda}$ and the
parameter $k_Fa$, which gives a measure of the Fermi surface volume or
alternatively the spacing of the chain. We consider only a
polarization in the $x$-$z$ plane. 
In the topologically non-trivial regions, we plot the minimum gap
magnitude, demonstrating the existence of a fully gapped state; the
non-topological regions are left white. The most important aspect of
this phase diagram is that a topological state is revealed to be
possible even for infinitesimal SOC. Remarkably, the excitation
  spectrum can display a substantial
  gap even for very small SOC strength $\widetilde{\lambda}\ll 1$. We
emphasize that our analysis is only valid for $\epsilon_0$ sufficiently
close to zero, and so other methods are required to comprehensively
survey the phase diagram. 

\section{Summary}

In this paper we have studied the appearance of a
topological impurity band when a ferromagnetic chain of classical
spins are embedded in a two-dimensional singlet $s$-wave
superconductor with Rashba SOC. To this end, we have
derived an effective tight-binding model for the overlapping YSR
states of the impurities. When the spins are polarized perpendicular
to the SOC along the chain, an odd-parity pairing term is induced in
the effective Hamiltonian, thus realizing a 
Kitaev-like model with generically non-trivial topology. Our
work and recent others~\cite{Heimes2014} explore
alternative routes to a topological YSR 
chain which do not rely upon helical spin
texture.~\cite{Choy2011,NadjPerge2013,Nakosai2013,Klinovaja2013,Braunecker2013,Vazifeh2013,Pientka2013,Poeyhoenen2014,Pientka2014,Reis2014,Roentynen2014,NadjPergeAPS2014}
This is 
a significant result, as the stability of the helical spin texture is
debated.~\cite{Reis2014,Kim2014} In contrast, the SOC
mechanism examined here is intrinsic to the superconductor
surface. This implies that topological phases
are possible for a much wider variety of impurity spin
configurations than hitherto realized, which grants the YSR chain
proposal additional robustness and lends strong theoretical 
support to experimental efforts to detect Majorana fermions in such a 
setting. As revealed by our calculated quantum phase
diagram~\fig{phasediagram}, however, the topological phase in the
ferromagnetic YSR chain system is not generic. Some
fine-tuning of the system is therefore required in order to observe
topological Majorana fermions through the measurement, for example, of
zero-bias-conductance peaks in tunneling spectroscopy experiments. 

Although we have confined ourselves to the analytically-tractable
limit of a dilute chain of deep impurities, we expect that our
results are of more general validity since they rely only upon the
low-energy form of the Green's function. We have also neglected
complicating factors such as particle-hole asymmetry in the normal
state dispersion, the
suppression of the superconducting gap close to the impurity
spins, and the three-dimensional nature of the superconducting
host. These issues must certainly be accounted for when modelling a
realistic system, but can only be addressed using large-scale
computer simulations. Nevertheless, none of these effects should
invalidate the mechanism giving rise to the topological state of our
basic model which arises simply from the interplay between
ferromagnetism, superconductivity, and SOC.  

\begin{acknowledgments}
The authors thank P. Kotetes
and A. Yazdani for useful discussions. This work is supported by
JQI-NSF-PFC and LPS-CMTC.
\end{acknowledgments}

\appendix

\section{Important integrals}

In this appendix we present analytic forms for the four
integrals~\eq{eq:impintegrals} encountered in our solution of the
YSR chain. We perform by these integrals
by extending the cutoff $D\rightarrow\infty$. We
distinguish two cases for the argument: $x=0$ for the isolated
YSR impurity, and $x\neq0$ for the YSR chain. 

\subsection{Isolated impurity: $x=0$}

In this case all the integrals except $I^{\nu}_3(0)$ are vanishing,
which evaluates to
\beq
I^{\nu}_3(0) = -\frac{\pi{\cal N}_\nu}{\sqrt{\Delta^2 -\omega^2}}\,.
\eeq

\subsection{Impurity chain: $x\neq0$}

For $x\neq0$, we first evaluate the integral over $\xi$ using
elementary contour integral methods, and then evaluate the angular
integral. We hence find
\begin{widetext}
\beqarray
I^\nu_{1}(x) & = & \pi{\cal N}_\nu\text{Im}\left\{J_{0}((k_{F,\nu} +
i\xi_{\nu}^{-1})|x|) + iH_{0}((k_{F,\nu} +
i\xi_{\nu}^{-1})|x|)\right\} \,, \\
I^\nu_2(x) &=& -i\pi{\cal N}_\nu\sgn(x)\text{Re}\left\{iJ_{1}((k_{F,\nu} +
i\xi_{\nu}^{-1})|x|) + H_{-1}((k_{F,\nu} +
i\xi_{\nu}^{-1})|x|)\right\} \,, \\
I^\nu_{3}(x) & = & -\frac{\pi{\cal N}_\nu}{\sqrt{\Delta^2 - \omega^2}}\text{Re}\left\{J_{0}((k_{F,\nu} +
i\xi_{\nu}^{-1})|x|) + iH_{0}((k_{F,\nu} +
i\xi_{\nu}^{-1})|x|)\right\} \,, \\
I^\nu_{4}(x) &=& -\sgn(x)\frac{i\pi{\cal N}_\nu}{\sqrt{\Delta^2 -
    \omega^2}}\text{Im}\left\{iJ_{1}((k_{F,\nu} +
i\xi_{\nu}^{-1})|x|) + H_{-1}((k_{F,\nu} +
i\xi_{\nu}^{-1})|x|)\right\} \,,
\eeqarray
where $J_n(z)$ and $H_n(z)$ are Bessel and Struve functions of order
$n$, respectively, and $\xi_{\nu} = \hbar
v_{F,\nu}/\sqrt{\Delta^2-\omega^2}$. Using asymptotic
forms~\cite{AbramowitzStegun} 
valid for large values of the argument close to the positive real
axis, we can approximate these as
\beqarray
I^\nu_{1}(x) & \approx & \pi{\cal N}_\nu\sqrt{\frac{2}{\pi k_{F,\nu}|x|}}\sin\left(k_{F,\nu}|x| -
\tfrac{\pi}{4}\right)e^{-|x|/\xi_{\nu}} +
\frac{2{\cal N}_\nu}{k_{F,\nu}|x|} \,, \\
I^\nu_{2}(x) & \approx & i\pi {\cal N}_\nu\sgn(x)\sqrt{\frac{2}{\pi k_{F,\nu} |x|}}\sin\left(k_{F,\nu}|x|
-\tfrac{3\pi}{4}\right)e^{-|x|/\xi_{\nu}}  +
\sgn(x)\frac{2i{\cal N}_\nu}{(k_{F,\nu} x)^2}\,,\\
I^\nu_{3}(x) & \approx & -\frac{\pi{\cal N}_\nu}{\sqrt{\Delta^2 - \omega^2}}\sqrt{\frac{2}{\pi k_{F,\nu}|x|}}\cos\left(k_{F,\nu}|x| -
\tfrac{\pi}{4}\right)e^{-|x|/\xi_{\nu}}\,,\\
I^\nu_{4}(x) & \approx & -\sgn(x)\frac{i\pi{\cal N}_\nu}{\sqrt{\Delta^2 -
    \omega^2}}\sqrt{\frac{2}{\pi k_{F,\nu}|x|}}\cos\left(k_{F,\nu}|x| -
\tfrac{3\pi}{4}\right)e^{-|x|/\xi_{\nu}}\,.
\eeqarray
The nonoscillating component is valid up to ${\cal O}((k_{F,\nu}|x|)^{-3})$.

\subsection{Fourier transforms}

The Fourier transform of the effective
Hamiltonian~\eq{eq:Heffrealspace} can be carried out analytically when
we utilize the asymptotic expressions. Defining the Fourier
transform as
\beq
A(k) = \sum_{j}A_{0j}e^{ikja}\,,
\eeq
we obtain
\beqarray
I^{\nu}_2(k) & = &  {\cal N}_\nu\sqrt{\frac{\pi}{2
    k_{F,\nu}a}}\left\{e^{-3\pi i/4}\text{Li}_{\tfrac{1}{2}}\left(e^{i(k_{F,\nu}a
    + ka) - a/\xi_{\nu}}\right) - e^{3\pi i/4}\text{Li}_{\tfrac{1}{2}}\left(e^{i(-k_{F,\nu}a
    + ka) - a/\xi_{\nu}}\right)\right. \notag \\
& & \left. - e^{-3\pi i/4}\text{Li}_{\tfrac{1}{2}}\left(e^{i(k_{F,\nu}a
    - ka) - a/\xi_{\nu}}\right) - e^{3\pi i/4}\text{Li}_{\tfrac{1}{2}}\left(e^{i(-k_{F,\nu}a
    - ka) - a/\xi_{\nu}}\right)\right\} \notag \\
&& + \frac{2i{\cal N}_\nu}{(k_{F,\nu}a)^2}\left\{\text{Li}_2\left(e^{ika}\right) - \text{Li}_2\left(e^{-ika}\right)\right\}\,,\\
I^{\nu}_{3}(k) & = & -\frac{{\cal N}_\nu}{\sqrt{\Delta^2 - \omega^2}}\sqrt{\frac{\pi}{2
    k_{F,\nu}a}}\left\{e^{-\pi i/4}\text{Li}_{\tfrac{1}{2}}\left(e^{i(k_{F,\nu}a
    + ka) - a/\xi_{\nu}}\right) + e^{\pi i/4}\text{Li}_{\tfrac{1}{2}}\left(e^{i(-k_{F,\nu}a
    + ka) - a/\xi_{\nu}}\right)\right. \notag \\
& & \left. + e^{-\pi i/4}\text{Li}_{\tfrac{1}{2}}\left(e^{i(k_{F,\nu}a
    - ka) - a/\xi_{\nu}}\right) + e^{\pi i/4}\text{Li}_{\tfrac{1}{2}}\left(e^{i(-k_{F,\nu}a
    - ka) - a/\xi_{\nu}}\right)\right\} \,,\\
I^{\nu}_{4}(k) & = & -\frac{ i {\cal N}_\nu}{\sqrt{\Delta^2 - \omega^2}}\sqrt{\frac{\pi}{2
    k_{F,\nu}a}}\left\{e^{-3\pi i/4}\text{Li}_{\tfrac{1}{2}}\left(e^{i(k_{F,\nu}a
    + ka) - a/\xi_{\nu}}\right) + e^{3\pi i/4}\text{Li}_{\tfrac{1}{2}}\left(e^{i(-k_{F,\nu}a
    + ka) - a/\xi_{\nu}}\right)\right. \notag \\
& & \left. - e^{-3\pi i/4}\text{Li}_{\tfrac{1}{2}}\left(e^{i(k_{F,\nu}a
    - ka) - a/\xi_{\nu}}\right) - e^{3\pi i/4}\text{Li}_{\tfrac{1}{2}}\left(e^{i(-k_{F,\nu}a
    - ka) - a/\xi_{\nu}}\right)\right\}\,,
\eeqarray
where $\text{Li}_{s}(z)$ is the polylogarithm of order $s$. 
\end{widetext}

\end{document}